# First-principles calculation of the temperature dependence of the optical response of bulk GaAs

Z. A. Ibrahim (1), A. I. Shkrebtii[*] (1), M. J. G. Lee (2), K. Vynck (2), T. Teatro (1),
W. Richter (3 and 4), T. Trepk (3 and 5), T. Zettler (5)

((1) Faculty of Science, University of Ontario Institute of Technology, Oshawa, Canada,
(2) Department of Physics, University of Toronto, Toronto, Canada,
(3) Institut für Festkörperphysik, Technische Universität Berlin, Berlin, Germany,
(4) Dipartimento di Fisica, Università di Roma Tor Vergata, Rome, Italy,
(5) LayTec GmbH, Berlin, Germany)

A novel approach has been developed to calculate the temperature dependence of the optical response of a semiconductor. The dielectric function is averaged over several thermally perturbed configurations that are extracted from molecular dynamic simulations. The calculated temperature dependence of the imaginary part of the dielectric function of GaAs is presented in the range from 0 to 700 K. This approach that explicitly takes into account lattice vibrations describes well the observed thermally-induced energy shifts and broadening of the dielectric function.

Optical techniques yield a wealth of information about the fundamental properties of materials and are widely used in leading-edge technologies in the semiconductor industry [1]. Methods include spectroscopic ellipsometry (SE), reflectance anisotropy spectrometry (RAS) and second harmonic generation (SHG) [2]. These noninvasive tools are widely used to characterize materials and their growth, and to monitor chemical reactions at surfaces and interfaces. They can be applied in harsh chemical, intense radiation and high temperature environments [1] and do not require ultra-high vacuum (UHV) conditions.

*In-situ* characterization and monitoring of high quality epitaxial growth and thin film deposition are in increasing demand. Temperature being one of the key optimizing factors influencing the growth, it is important to establish a theoretical framework to relate the optical response to temperature dependent dynamical processes on the atomic scale.

An important step towards first-principles quantum mechanical calculations of the linear bulk and surface optical properties of semiconductors was taken in 1998, when excitonic effects with local field corrections were incorporated into the theory [3--5]. However, one of the key factors, namely atomic vibration, was neglected in these as well as in other calculations, which were all based on an unperturbed crystallographic lattice. To bring the sharp calculated peaks into better agreement with the experimental peaks, large Lorentzian broadening was used (see, e.g., [4]). To date, no optical response calculations are available that explicitly take into account temperature effects.

In this Letter we extend the optical response calculations to take into account the effect of temperature, and we illustrate our method by applying it to the imaginary part of the dielectric function $\varepsilon_2$ *(E)* of bulk GaAs. The physical mechanisms responsible for the temperature dependence of the optical properties are traditionally separated into lattice dynamics in the harmonic approximation (*i.e.*, phonons) and anharmonic effects including thermal expansion. Lattice vibrations can be associated with a temperature dependent mean square displacement (MSD), which is an important characteristic of the material. The phonon density of states (DOS) of bulk GaAs has been calculated by Giannozi *et al.* [6]. The MSD in the harmonic approximation calculated from the phonon DOS shows that at 0 K the MSD is governed by zero-point motion of the lattice. The calculated MSD can be used to extrapolate the relationship between MSD and temperature to the range 0 K to 80 K, where to our knowledge no experimental MSD have been reported. In the range 80 K to 600 K, the calculated MSD is in good agreement with experimental X-ray diffraction data [7] (original data from [8]). Above about 600 K the experimental data show that the MSD is enhanced by anharmonic effects.

Measurements of the dielectric function of GaAs in the temperature range from 22 K to 750 K have been reported by Cardona and coworkers [9]. We extend the dielectric function measurements to 1070 K. Our classical spectroscopic ellipsometer in a rotating polarizer-sample-analyzer (RPSA) configuration [10] is attached to a horizontal metal organic vapor phase epitaxy (MOVPE) reactor [11]. After chemically cleaning and etching the epiready GaAs (001) wafers, they were heated under group V-stabilization using arsine (*AsH₃*). The remaining oxide desorbted in the reactor's hydrogen atmosphere (10 kPa). Details for the *in-situ* monitoring of the oxide-desorption can be found in [12]. A homo-epitaxial buffer of more than 100 nm was grown at 923 K to overgrow any remaining nonidealities of the substrates [13].

The dielectric function was measured over a range of temperature settings of the heating system, which was calibrated by means of a eutectic sample. To avoid the increasing desorption of As that occurs at higher temperatures, the sample was heated to a maximum of 1070 K. The critical points of $\varepsilon_2$ *(E)* were determined from the second derivative in the parabolic band approximation [14,15].

Lattice vibrations modulate the overlap of the electronic wavefunctions between neighboring atoms, thereby influencing the widths of the energy bands, the energies of interband transitions and the optical response. In accordance with the ergodic hypothesis, the optical response of a bulk crystal is equivalent to the average over a sufficient number of thermal configurations of a small supercell. Because the time scale of electron transitions is much shorter than that of lattice vibrations, the optical response of GaAs was determined by averaging the optical response calculated adiabatically over several representative configurations of a thermally-perturbed cubic supercell of eight atoms (four Ga and four As).

The representative thermal configurations were extracted from first-principles molecular dynamic (MD) simulations based on density functional theory (DFT) using the software package Quantum-Espresso [16]. A plane wave basis set with an energy cutoff of 15 Ry was used, and the Brillouin-zone was sampled at 8 *k*-vectors. The equilibrium volume of the unperturbed supercell was determined by minimizing the total energy in the zinc-blende structure. Initially, the atoms of the supercell were randomly displaced to set the average temperature in the MD simulation. We found that 10000 steps, each of 20 atomic units of time (1 a.u. = 2.42 × $10^{-17}$ s), were sufficient to equilibrate the system, and we extracted the thermally-induced atomic configurations over the following 5000 steps. We found that it suffices to average the calculated $\varepsilon_2$ *(E)* over no more than eight thermal configurations distributed throughout the equilibrated part of the MD simulation. This was verified by comparing the results with tight-binding calculations of $\varepsilon_2$ *(E)* using up to one hundred configurations.

The effect of using a small supercell is to displace the phonon modes towards the Brillouin zone boundary. As a result the internal energy of the supercell overestimates the temperature of the thermal configuration. Therefore, we extracted the temperature from the MSD of each thermal configuration using the results of Giannozzi *et al.* [7]. Since the temperature fluctuates in the MD simulation, we selected only configurations within ±2.5 K of the average temperature below 200 K, within ±10 K in the range 200 K to 600 K, and within ±25 K above 600 K.

The software package WIEN2K [17] was used to calculate the electronic structure and the optical properties of each thermal configuration self-consistently on the basis of DFT by the full-potential linear-augmented-plane-wave (FP-LAPW) method. Exchange and correlation were treated within the generalized-gradient approximation (GGA) [18]. $\varepsilon_2 (E)$ was calculated within the random phase approximation (RPA) and a scissors shift of 0.8 eV was applied to correct the energies of the conduction states, which are known to be underestimated by DFT [19]. No Lorentzian broadening was applied. Because of the broken symmetry of the thermal configurations, a large number of **k**-vectors are needed to sample the Brillouin zone. 864 irreducible **k**-vectors were used in the self-consistent cycles, and 23300 irreducible **k**-vectors were used to calculate the optical matrix elements. We took into account the experimental temperature dependence of the lattice constant of bulk GaAs [20].

The periodic boundary conditions in our electronic structure calculations result in the replication of the supercell. As a consequence, some of the thermal configurations that are generated by our MD simulations show sizeable displacements of one or more low-index planes of atoms, resulting in unphysical band splitting. To eliminate these unrepresentative configurations, we used only configurations whose geometrical structure factors for the low-index planes {200}, {220} and {111} (hence {222}) are minimally perturbed compared with those of the unperturbed lattice.

Several interband transition peaks are clearly resolved in the calculated $\varepsilon_2 (E)$ at low temperatures below 200 K, as shown in Fig. 1. They include the direct gap transition onset $E_0$ close to 1.4 eV, peak $E_1$ at approximately 3.0 eV, the strongest peak $E_2$ around 5.0 eV with shoulder $E_0'$ close to 4.6 eV, and the weak peak $E_1'$ at 5.8 eV. At 0 K, due to zero-point motion, all of these peaks are significantly broadened, reduced in amplitude by about 25% and shifted to lower energy by about 50 meV as compared with the corresponding peaks calculated for the unperturbed GaAs lattice. With increasing temperature they are progressively broadened and shifted to lower energy. At 314 K and above, the shoulder $E_0'$ and peak $E_1'$ are smeared out, while peaks $E_1$ and $E_2$ remain strong. Bulk thermal expansion has only a minor effect on the calculated critical point energies of GaAs (linear shift of −3.6 meV per 100 K).

Our $\varepsilon_2 (E)$ measurements in the range 300 K to 1070 K are shown in Fig. 2. Measurements in the range 22 K to 750 K were reported by Cardona *et al.* [9]. With increasing temperature the principal optical peaks broaden, decrease in amplitude and shift to lower energy.

In Fig. 3, the calculated energy dependencies of peaks $E_1$ and $E_2$ in the temperature range from 0 K to 700 K are compared with the experimental data. The energy shifts of the calculated $\varepsilon_2(E)$ agree well with the experiments. In addition, the thermally induced broadening of the calculated $\varepsilon_2(E)$ is in good overall agreement with experiment.

To make the present computationally-intensive calculations more tractable, we did not take into account either excitonic effects or the spin-orbit interaction. Excitonic effects are known to enhance peak $E_1$ and to suppress peak $E_2$ in GaAs [4], but they are not expected to influence the relative energies of these peaks [21]. The neglect of excitonic effects in our calculations results in the discrepancy in the relative strengths of peaks $E_1$ and $E_2$ between our calculations and experiments. Peaks $E_1$ and $E_1+\Delta_1$, which are clearly resolved in the experimental data below 300K (see Fig. 2), appear as a single peak in our calculated $\varepsilon_2 (E)$ due to the neglect of the spin-orbit interaction.

In conclusion, we proposed an approach that explicitly takes into account lattice dynamical effects in the calculation of the linear optical response of semiconductors. The dielectric function at finite temperature is deduced by averaging the dielectric functions over various thermal configurations that are extracted from first-principles molecular dynamic simulations. This, for the first time, allowed to accurately account for the thermally-induced energy shift and broadening of $\varepsilon_2 (E)$, including those observed at 0 K that are caused by zero-point motion. In light of recent developments in including excitonic effects in the theory of the optical response, it now becomes feasible to combine a treatment of excitonic and spin-orbit effects with this explicit treatment of lattice dynamics to obtain a comprehensive first-principles description of the optical properties at finite temperature.

## Acknowledgments

M. Cardona's interest and comments are highly appreciated. We wish to acknowledge partial financial support from the Natural Sciences and Engineering Research Council of Canada (NSERC), SHARCNET and Sfb 296. We thank D. Zekveld for his participation in the initial stage of the project.

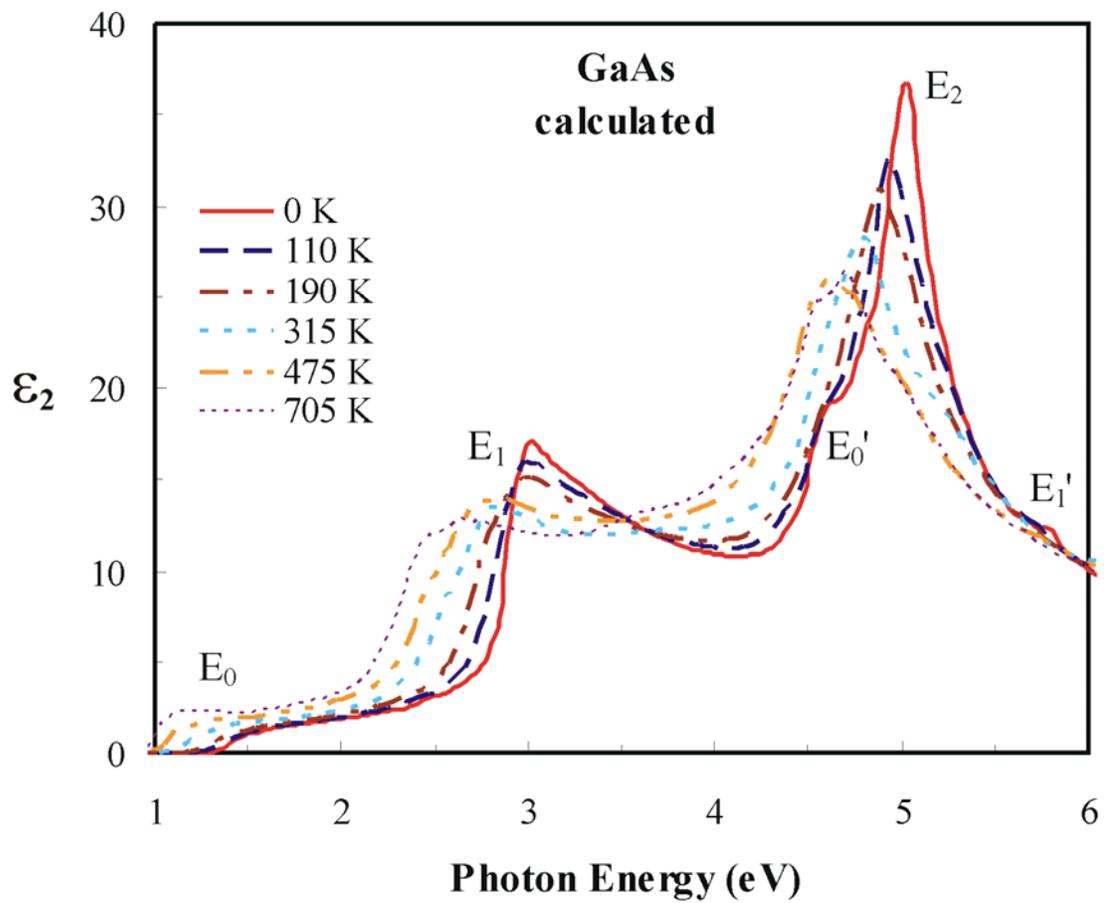

FIG. 1 (color online). Calculated imaginary part of the dielectric function of GaAs in the range from 0 K to 700 K.

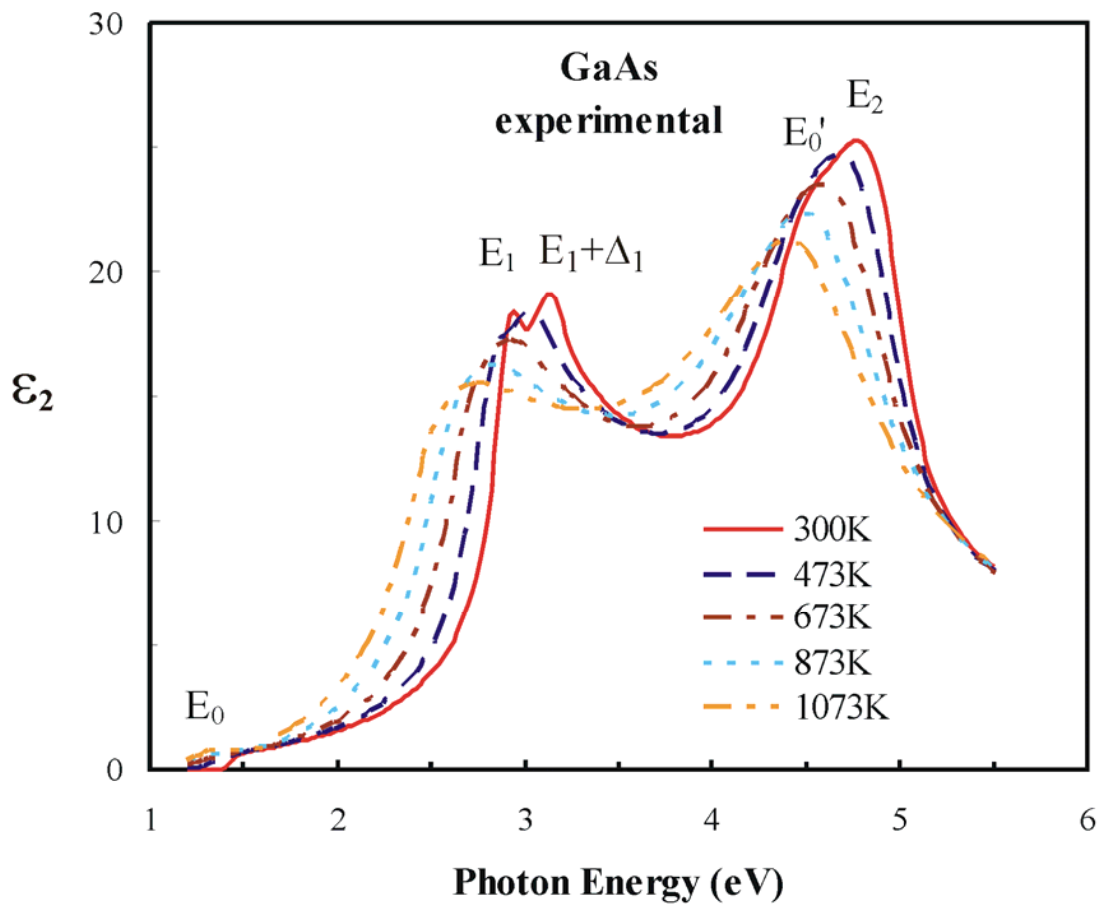

FIG. 2 (color online). Measured imaginary part of the dielectric function of GaAs in the range from 300 K to 1070 K.

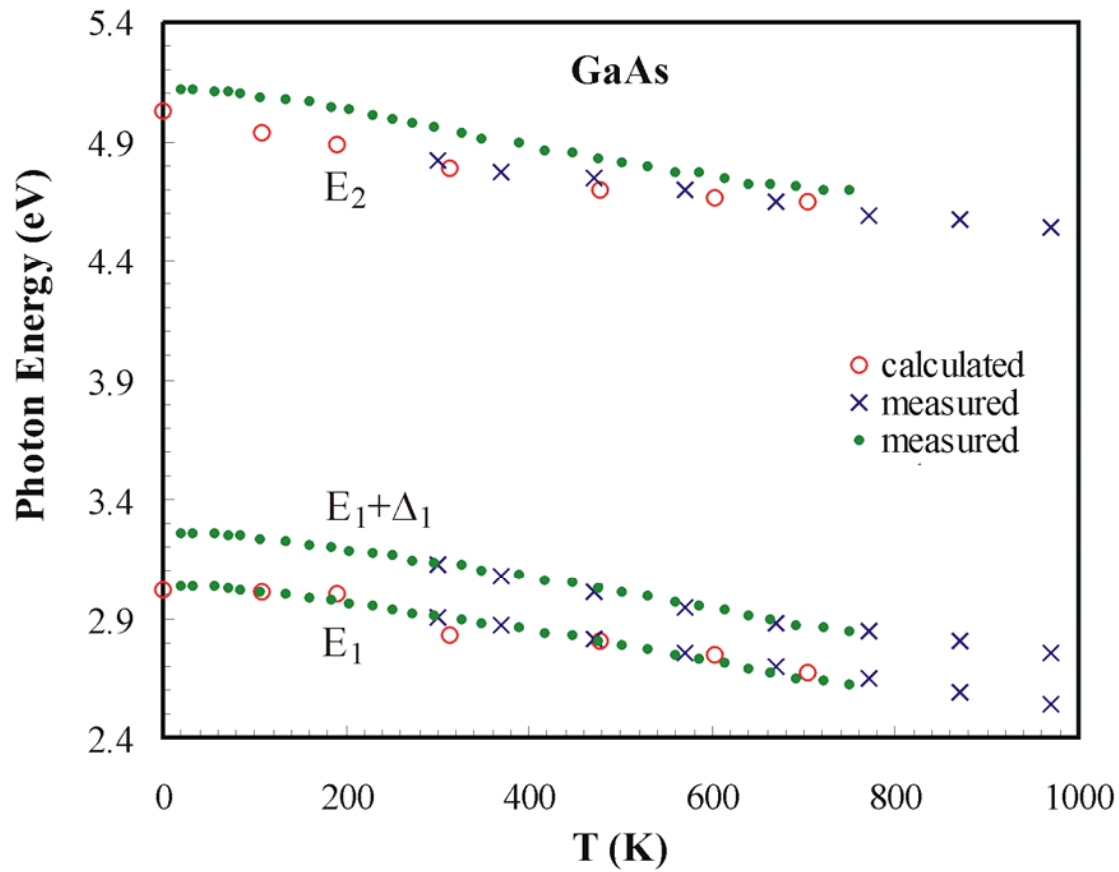

FIG. 3 (color online). Temperature dependence of the main critical point energies of the imaginary part of the dielectric function of GaAs. Calculated data are represented by circles; measured data by crosses and by dots (Ref. [9]).

9